\documentclass[11 pt]{article}

\usepackage{amsmath,amssymb}

\begin{document}

\title{\LARGE{   \bf{Soliton Solutions with Real Poles in the Alekseev formulation of the Inverse-Scattering method}}}
 
\author{
        {\bf{S. Miccich\`e}}  {\footnote{   {  {\it{e-mail}:}~{\tt S.Micciche@Lboro.ac.uk  }  }  }  }
              {\rm{~and~}}
        {\bf{J.B. Griffiths}} {\footnote{   {  {\it{e-mail}:}~{\tt J.B.Griffiths@Lboro.ac.uk  }  }  }  } \\
        {\small{\it{{Department of Mathematical Sciences, Loughborough University,}}}} \\
        {\small{\it{{Loughborough, Leics., LE11 3TU (GB)}}}}
       }

\date{\today}

\maketitle


\def\Tr{\mathop{\rm tr}\limits}

\def\Re{\mathop{\rm Re}\limits}

\def\Im{\mathop{\rm Im}\limits}



\begin{abstract}

A new approach to the inverse-scattering technique of Alekseev is
presented which permits real-pole soliton solutions of the Ernst equations to be considered. This
is achieved by adopting distinct real poles in the scattering matrix and its
inverse. For the case in which the electromagnetic field vanishes, some
explicit solutions are given using a Minkowski seed metric. The relation with
the corresponding soliton solutions that can be constructed using the
Belinskii--Zakharov inverse-scattering technique is determined.

\end{abstract}

\section{Introduction}  \label{intro}

The Einstein--Maxwell field equations for a space-time endowed with two nonnull, commuting Killing vectors are exactly integrable. Over the past few decades a number of techniques have been developed by which exact solutions of these equations may be constructed. In this paper we will consider a modification of the soliton technique introduced by Alekseev \cite{A2,A1}.

The inverse-scattering technique of Belinskii and Zakharov (BZ) is now well
known \cite{BZ1,BZ2}. It is a solution-generating procedure for producing exact vacuum
solutions. Starting from some initial ``seed'' solution, the technique is
based on the construction of a scattering, or ``dressing'' matrix which is a
meromorphic function of a complex spectral parameter. It is
essentially modelled on the usual inverse-scattering methods for solving
nonlinear p.d.e.'s such as the Korteweg--de~Vries or Sine--Gordon equations.  The
alternative inverse-scattering
technique of Alekseev similarly generates solutions of the
Einstein--Maxwell equations (for specific examples see \cite{A1,DGN1,DaGlNi93,GNG1,GG1,GT}). In the case in which the
electromagnetic field vanishes, these electromagnetic solitons become
purely gravitational and may be equivalent to those that can be obtained
using the BZ technique.

For complex poles, the vacuum $N$-pole solitons in the Alekseev formalism are
equivalent to $2N$-pole solitons in the BZ formalism \cite{DGN1,GG1,DGGP,KO2}, although it may be noted that different spectral planes are adopted in the two formalisms. In order for the metric to be
real, it is necessary that any complex pole in the BZ approach must be
accompanied by its complex conjugate as another pole. In the Alekseev approach a single complex pole is permitted, but a
distinct (normally complex conjugate) pole must appear in the inverse of the scattering matrix. However, in the BZ approach, a single real pole is
permitted and this is ruled out in the
standard Alekseev approach. The reason for this is that it would lead to the same pole in the
inverse matrix and, in this case, the subsequent procedure becomes singular.

These simple arguments seem to show that Alekseev's class of vacuum-solitons is
smaller than that of Belinskii--Zakharov, since real poles are not
permitted. It is the purpose of this paper to present a new approach to the
construction of real pole solitons in the Alekseev formalism. This will be
achieved by introducing distinct real poles in the inverse of the scattering matrix. In section \ref{sec:explsol} we will present explicit solutions for the case of one real pole and a Minkowski seed. This will indicate that the solutions generated by this method are equivalent to BZ soliton solutions with multiple real poles.

\section{Alekseev's Soliton Technique}   \label{SolitonTechnique}

Let us consider spacetimes with two commuting Killing vectors:
\begin{subequations}  \label{2KV}
\begin{eqnarray}
	     &&  ds^2 = -f~\eta_{_{AB}}~d x^{_{A}} d x^{_{B}} - g_{_{ab}} d x^a d x^b  
                     \qquad   
                 \eta_{_{AB}} = \left( \begin{array}{cc}
                                                         - \epsilon & 0 \\
                                                         0          & 1
                                \end{array} \right)  
                 \qquad   
                 A = 1,2 ~~ a = 3,4 \\
             && \nonumber \\
             && x^{_A} = (\alpha, \beta) 
                          \qquad
                 f = f(\alpha, \beta) 
                           \qquad
                 g_{_{ab}} = g_{_{ab}}(\alpha, \beta) 
                           \qquad
                 \det g_{_{ab}} = \epsilon~\alpha^2 
                           \qquad 
                 \epsilon = \pm 1
\end{eqnarray}
\end{subequations}
where the $x^{_a}$'s are ignorable coordinates. For these spacetimes, the Ernst equations for electrovacuum solutions are the integrability conditions for Alekseev's linear pair of equations \cite{A2,A1}:
\begin{eqnarray}
                \partial_{_{A}}~{\boldsymbol \Psi} = \Lambda_{_{A}}^{\ ^{_B}}~{\bf{U}}_{_{B}}~{\boldsymbol{\Psi}},
                \qquad \qquad 
                \Lambda_{_{A}}^{\ ^{_B}}~=~{1 \over {2 i}}~
                                        {  { (w - \beta)~\delta_{_{A}}^{\ ^{_B}}~+~\epsilon~\alpha~\epsilon_{_{A}}^{\ _{B}}  } 
                                             \over 
                                           { (w - \beta)^2~-~\epsilon~\alpha^2 }
                                        }  ,                    \label{12.13}        
\end{eqnarray}
where $w$ is the (unphysical) spectral parameter. The $3 \times 3$ matrices ${\bf{U}}_{_{A}}$ are related to the metric components $g_{_{ab}}$ and the electromagnetic components $\Phi^a$ through the so-called ``additional conditions'':
\begin{subequations} \label{11.48}
\begin{eqnarray}
               && \partial_{_{A}}~(~{\bf{G}}~ - 4~i~\beta~{\boldsymbol{\Omega}}~)~=~2~\bigl(~{{\bf{U}}^\dagger}_{_{A}}~{\boldsymbol{\Omega}}~-~{\boldsymbol{\Omega}}~{\bf{U}}_{_{A}}~\bigl),  \label{11.48.a}  \\
               && {\bf{G}}~{\bf{U}}_{_{A}}~=~-~4~i~\epsilon~\alpha~\epsilon_{_{A}}^{\ _B}~{\boldsymbol{\Omega}}~{\bf{U}}_{_{B}},   \label{11.48.b}  \\ 
               &&  \Tr {\bf{U}}_{_{A}}~=~ 2 i~\epsilon~\epsilon_{_{A}}^{\ _{B}}~\partial_{_{B}} \alpha ~,
                     \qquad  \qquad
                   {\rm{Re}} \bigl ( \Tr {\bf{U}}_{_{A}} \bigl ) =~0~,   \label{11.48.c}                    
\end{eqnarray}
\end{subequations}
with $\epsilon_{_{A}}^{\ _B} = \eta_{_{AC}}\,\epsilon^{_{CB}}$. Also 
\begin{eqnarray}
                  {\bf{G}} = \left( \begin{array}{cc}
                                          - 4 h^{ab} + 4 {\Phi}^a \overline{{\Phi}}^b & - 2 {\Phi}^a  \\
                                                                                   &             \\
                                          - 2 \overline{\Phi}^b                    & 1           
                      \end{array} \right),
                         ~~
                  {\boldsymbol{\Omega}} = \left ( \begin{array}{ccc}
                                              0 & 1 & 0  \\
                                             -1 & 0 & 0  \\
                                              0 & 0 & 0  \
                     \end{array} \right),
                        ~~
                  \epsilon_{_{AB}} = \epsilon^{_{AB}} = \left( \begin{array}{cc}
                                                                        0 & 1 \\
                                                                       -1 & 0 
                                                          \end{array}   \right)  ,  \label{ALEKbricks}
\end{eqnarray}
$h^{ab} = \epsilon^{ac}\,\epsilon^{bd}\,g_{cd}$ and a $\dagger$ denotes the hermitian conjugate. It is also convenient to introduce the matrix function ${\bf{W}} = {\bf{G}} + 4 i\,(w - \beta)\,{\boldsymbol{\Omega}}$, which is linear in $w$. It can then be shown that the linear equations $(\ref{12.13})$ and the above conditions imply that the matrix ${\bf{K}} \equiv {\boldsymbol{\Psi}}^\dagger~{\bf{W}}~{\boldsymbol{\Psi}}$ depends on the  spectral parameter only. This first integral of the linear equations plays a central role in the solution-generating procedure.

Given any seed metric ${\bf{g}}_{_{0}}$ and the associated matrix ${\boldsymbol{\Psi}}_{_{0}} (\alpha, \beta,w)$, a new solution for the linear pair is given by a dressing ansatz:
\begin{eqnarray}
               {\boldsymbol{\Psi}}(\alpha, \beta, w)~=~{\boldsymbol{\chi}}(\alpha, \beta, w)~{\boldsymbol{\Psi}}_{_{0}} (\alpha, \beta,w)        \label{12.32} 
\end{eqnarray}
The dressing matrix ${\boldsymbol{\chi}}$ and its inverse are assumed to have meromorphic structure with respect to the spectral parameter $w$. Specifically:
\begin{subequations}  \label{13}
\begin{eqnarray}
              &&  {\boldsymbol{\chi}}     ~=~{\bf{I}}~+~\sum_{k = 1}^{N}~{1 \over{w -   w_{_{k}}}}~{\bf{R}}_{_{k}}    \qquad  \quad
                  {\boldsymbol{\chi}}^{-1}~=~{\bf{I}}~+~\sum_{k = 1}^{N}~{1 \over{w - \tilde{w}_{_{k}}}}~{\bf{S}}_{_{k}}    \label{13.1} \\
              && {\bf{R}}_{_{k}}~=~{\bf{n}}_{_{k}}~\otimes~{\bf{m}}_{_{k}} 
                 \qquad \qquad  \quad \quad ~~~~
                 {\bf{S}}_{_{k}}~=~{\bf{p}}_{_{k}}~\otimes~{\bf{q}}_{_{k}}   \label{13.2} 
\end{eqnarray}
\end{subequations}
Equations $(\ref{13.2})$ take into account the fact that both ${\bf{R}}_{_{k}}$ and ${\bf{S}}_{_{k}}$ must be matrices with vanishing determinant. The new metric ${\bf g}$ and the new electromagnetic field components $\Phi^a$ may be determined as the components of the matrix ${\bf{G}}$ given by:
\begin{eqnarray}
               {\bf{G}}~=~{\bf{G}}_{_{0}}~-~4 i~\Bigl( {\boldsymbol{\Omega}}~{\bf{R}}~+~{\bf{R}}^\dagger~{\boldsymbol{\Omega}} \Bigl)~+~4 i~\beta_{_{0}}~{\boldsymbol{\Omega}}~,
                           \qquad \qquad
                {\bf{R}} =\sum_{k = 1}^N~{\bf{R}}_{_{k}}~.  \label{12.46}
\end{eqnarray}
In fact, it is still necessary to impose the condition that the new metric component $h_{_{12}}$ is real. If the imaginary part turns out to be a non-zero constant, this can be removed by a suitable choice of the constant $\beta_{_{0}}$. With $(\ref{12.46})$ we have:
\begin{eqnarray}
                {\bf{W}}~={\bf{W}}_{_{0}}~-~4 i~\Bigl( {\boldsymbol{\Omega}}~{\bf{R}}~+~{\bf{R}}^\dagger~{\boldsymbol{\Omega}} \Bigl)~+~4 i~\beta_{_{0}}~{\boldsymbol{\Omega}}       \label{12.47}
\end{eqnarray}
Unlike the BZ case, the poles $w = w_{_{k}}$ and $w = \tilde{w}_{_{k}}$ are complex constants: they do not depend upon $\alpha$ and $\beta$. Finally, the vectors ${\bf{n}}_{_{k}}$, ${\bf{m}}_{_{k}}$, ${\bf{p}}_{_{k}}$ and ${\bf{q}}_{_{k}}$ are given in terms of arbitrary constant vectors ${\bf{k}}_{_{k}}$ and ${\bf{l}}_{_{k}}$ by:
\begin{subequations}   \label{enneandq}
\begin{eqnarray}
               && {\bf{q}}_{_{k}} = - \sum_{j=1}^N~\Gamma_{_{jk}}^{-1}~{\bf{m}}_{_{j}}    
                       \qquad 
                  {\bf{n}}_{_{k}} = \sum_{j=1}^N~\Gamma_{_{kj}}^{-1}~{\bf{p}}_{_{j}}    
                        \qquad 
                  \Gamma_{_{kj}} = {  {{\bf{p}}_{_{k}} \cdot {\bf{m}}_{_{j}}} \over {w_{_{j}} - \tilde{w}_{_{k}}} }    \label{13.7}    \\
               && {\bf{m}}_{_{k}}~=~{\bf{k}}_{_{k}} \cdot {{\boldsymbol{\Psi}}_{_{0}}}^{-1}(w_{_{k}})    
                       \qquad 
                  {\bf{p}}_{_{k}}~=~{{\boldsymbol{\Psi}}_{_{0}}}~(\tilde{w}_{_{k}}) \cdot {\bf{l}}_{_{k}}           \label{13.14}
\end{eqnarray}
\end{subequations}

Let us now consider the conditions:
\begin{eqnarray}
             {{\boldsymbol{\Psi}}_{_{0}}}^{\dagger}~{\bf{W}}_{_{0}}~{\boldsymbol{\Psi}}_{_{0}}~=~{\bf{K}}_{_{0}}(w)~,
                      \qquad \qquad
             {{\boldsymbol{\Psi}}_{_{0}}}^{\dagger}~{\boldsymbol{\chi}}^\dagger~{\bf{W}}~{\boldsymbol{\chi}}~{\boldsymbol{\Psi}}_{_{0}}~=~{\bf{K}}(w)~.     \label{12.37}     
\end{eqnarray}
If, for simplicity, we choose ${\bf{K}}(w) = {\bf{K}}_{_{0}}(w)$, we obtain:
\begin{eqnarray}
                {\boldsymbol{\chi}}^{\dagger}~{\bf{W}}~{\boldsymbol{\chi}}=~{\bf{W}}_{_{0}}.           \label{12.38}
\end{eqnarray}
As we will show in the next section, the above assumption yields the condition $\tilde{w}_{_{k}} = \overline{w}_{_{k}}$. Indeed, this is the choice made by Alekseev $\cite{A2,A1}$. It is worth mentioning that this choice is not compulsory. In fact, there exists no issue in the construction of the soliton solutions outlined  above, such that one is forced to perform such a step. 

The above constraint immediately highlights the problems mentioned in $\S~\ref{intro}$. Let us consider $N=1$, so that the function $\Gamma$ in $(\ref{13.7})$ is simply given by:
\begin{eqnarray}
                  \Gamma  = {  {{\bf{p}}_{_{1}} \cdot {\bf{m}}_{_{1}}} \over {w_{_{1}} - \overline{w}_{_{1}}} }~.  \label{gamma1solst}
\end{eqnarray}
The denominator of $\Gamma$ is the difference between the poles $w_{_{1}}$ and $ \tilde{w}_{_{1}} = \overline{w}_{_{1}}$. These are the distinct (complex conjugate) poles in the dressing matrix ${\boldsymbol{\chi}}$ and in its inverse ${\boldsymbol{\chi}}^{-1}$. As long as $w_{_{1}}$ is a true complex number, $\Gamma$ is well defined and   it is possible to construct the appropriate $1$-soliton solution. However, if $w_{_{1}}$ is real, the denominator of $\Gamma$ vanishes and it is no longer possible to construct any solution. A similar problem occurs in the case of $N$ poles if any of these is real.

\section{A Generalized Construction for Real Poles Solitons}  \label{RPvac}

As stated above, a new solution of Einstein's Field Equations is given in terms of a previously-known solution by introducing the dressing ansatz $(\ref{12.32})$. Equations $(\ref{13})$ and $(\ref{enneandq})$ provide a new explicit $N$-pole solution. In order to obtain $(\ref{13.7})$ and $(\ref{13.14})$, no assumption has been made on the nature of the poles $w_{_{k}}$  in ${\boldsymbol{\chi}}$ and $\tilde{w}_{_{k}}$ in ${\boldsymbol{\chi}}^{-1}$. It is only assumed that they are simple. In the standard Alekseev technique $\cite{A1}$ it is assumed that ${\bf{K}}(w) = {\bf{K}}_{_{0}}(w)$ in $(\ref{12.37})$, but we wish to relax this assumption here. 

Let us introduce the matrix function $\tilde{\bf{W}}$ defined as:
\begin{eqnarray}
                \tilde{\bf{W}}~=~{\boldsymbol{\chi}}^\dagger~{\bf{W}}~{\boldsymbol{\chi}},~~~~{\rm{so~that}}~~~~
                {\bf{K}}(w)~=~{\boldsymbol{\Psi}}_{_{0}}^\dagger~\tilde{\bf{W}}~{\boldsymbol{\Psi}}_{_{0}}. \label{Wtilde} 
\end{eqnarray}
Obviously, if we choose ${\bf{K}} = {\bf{K}}_{_{0}}$, then $\tilde{\bf{W}} = {\bf{W}}_{_{0}}$ and the above definition coincides with $(\ref{12.38})$. Rewriting $(\ref{Wtilde})$ in the form $\tilde{\bf{W}}~{\boldsymbol{\chi}}^{-1} = {\boldsymbol{\chi}}^\dagger~{\bf{W}}$ gives
\begin{eqnarray}
		\tilde{\bf{W}} + \sum_k {1 \over {w - \tilde{w}_{_{k}}} }~\tilde{\bf{W}}~{\bf{S}}_{_{k}}~=~
                       {\bf{W}} ~+~\sum_k {1 \over {w - \overline{w}_{_{k}}} }~{{\bf{R}}_{_{k}}}^\dagger~{\bf{W}}    \label{important}
\end{eqnarray} 
If ${\bf{K}} = {\bf{K}}_{_{0}}$ then both ${\bf{W}}$ and $\tilde{\bf{W}}$ are linear in $w$. In order to have the same pole structure in both members of the above equation, it is then necessary to set $\tilde{w}_{_{k}} = \overline{w}_{_{k}}$. However, if we relax the initial assumption, then $\tilde{\bf{W}}$  is not linear in $w$ and may contain poles. We might thus be able to retain $\tilde{w}_{_{k}} \neq \overline{w}_{_{k}}$. In this section we will show that this second approach yields a positive answer to the question posed in $\S~\ref{intro}$.

\subsection{The $1$-Soliton Solution}  \label{1solsolution}

Let us consider the dressing matrices $(\ref{13})$ for a single real pole:
\begin{subequations} \label{1solRP}
\begin{eqnarray}
		&& {\boldsymbol{\chi}} = {\bf{I}} + { 1 \over {w - w_{_{1}} } }~{\bf{R}}
                  \qquad  \qquad \qquad 
                  {\boldsymbol{\chi}}^{-1} = {\bf{I}} + { 1 \over {w - \tilde{w}_{_{1}}} }~{\bf{S}}   \label{1solRPa}\\
                 && \nonumber \\
                &&  {\bf{R}} = {\bf{n}} \otimes {\bf{m}}
                  \qquad    
                   {\bf{S}} = {\bf{p}} \otimes {\bf{q}}
                   \qquad  \quad
                  w_{_{1}} , \tilde{w}_{_{1}}  \in {\mathbb R}
                   \qquad  
                  w_{_{1}}  \neq  \tilde{w}_{_{1}}  
\end{eqnarray} 
\end{subequations}
where ${\bf{n}}$, ${\bf{m}}$, ${\bf{p}}$, ${\bf{q}}$ are given by $(\ref{enneandq})$ and the suffix ${\scriptstyle{1}}$ has mostly been removed. 

In order to show that such a solution exists, it is necessary to prove that the matrix ${\boldsymbol{\Psi}}^{\dagger}~{\bf{W}}~{\boldsymbol{\Psi}}$ is a function of the spectral parameter only. To this purpose, let us consider the matrix function $\tilde{\bf{W}}$ defined above. With $(\ref{1solRPa})$, equation $(\ref{Wtilde})$ can be displayed explicitly as:
\begin{eqnarray}
              \tilde{\bf{W}}~=~{\bf{W}}~+~{1 \over {w- w_{_{1}}}}~\Bigl[
                                                        {\bf{R}}^\dagger~{\bf{W}}~+~{\bf{W}}~{\bf{R}}
                                                  \Bigl]~+~
                      {1 \over {(w- w_{_{1}})^2}}~{\bf{R}}^\dagger~{\bf{W}}~{\bf{R}} .              \label{expWtil1}
\end{eqnarray}
Writing ${\bf{W}} = {\bf{W}}(w_{_{1}}) + 4 i~(w - w_{_{1}})~{\boldsymbol{\Omega}}$, and using $(\ref{12.47})$ we obtain:
\begin{eqnarray}
           &&  \tilde{\bf{W}} = {\bf{W}}_{_{0}} + 4 i\,\beta_{_{0}}\,{\boldsymbol{\Omega}} +   
                            {1 \over {w- w_{_{1}}}}\>\Bigl[
                                                    {\bf{R}}^\dagger\,{\bf{W}}(w_{_{1}}) + {\bf{W}}(w_{_{1}})\,{\bf{R}} + 4 i\,{\bf{R}}^\dagger\,{\boldsymbol{\Omega}}\,{\bf{R}}
                                              \Bigl]    \nonumber \\
             &&  \hspace{1.3 in}   
                             +   {1 \over {(w- w_{_{1}})^2}}\>{\bf{R}}^\dagger\,{\bf{W}}(w_{_{1}})\,{\bf{R}} .  \label{expWtilde}
\end{eqnarray} 
We also note that, for a single pole, equation $(\ref{important})$ becomes:
\begin{eqnarray}
		\tilde{\bf{W}} +  {1 \over {w - \tilde{w}_{_{1}} } }~\tilde{\bf{W}}~{\bf{S}}~=~
                       {\bf{W}} ~+~ {1 \over {w- w_{_{1}}}}~{\bf{R}}^\dagger~{\bf{W}}  . \label{important1sol}
\end{eqnarray}
Since ${\bf{W}}$ is linear in $w$, equation $(\ref{important1sol})$ indicates that $\tilde{\bf{W}}$ must have a simple pole at $w = w_{_{1}}$. This implies that ${\bf{R}}^\dagger~{\bf{W}}(w_{_{1}})~{\bf{R}} = 0$ and hence:
\begin{eqnarray}
               \overline{\bf{n}} \cdot {\bf{W}}(w_{_{1}}) \cdot {\bf{n}}~=0~. \label{first constr} 
\end{eqnarray}
From equation $(\ref{important1sol})$, we also require that $\tilde{\bf{W}}(\tilde{w}_{_{1}})~{\bf{S}} = 0$, and hence:
\begin{eqnarray}
               \tilde{\bf{W}}(\tilde{w}_{_{1}}) \cdot {\bf{p}} = 0~. \label{sec constr} 
\end{eqnarray}
An  exact $1$-soliton solution will exist provided the two constraints $(\ref{first constr})$ and $(\ref{sec constr})$ are satisfied and provided the expression ${\boldsymbol{\Psi}}_{_{0}}^\dagger~\tilde{\bf{W}}~{\boldsymbol{\Psi}}_{_{0}}$, using $(\ref{expWtilde})$, depends on the spectral parameter only. In fact the latter constraint will be satisfied only for very particular seed metrics.

\bigskip

\noindent {\bf{3.1.1 An ansatz for the vacuum case}}.

\medskip

To generate a vacuum solution, it is necessary to start with a vacuum seed and  to put $k_{_{3}}=l_{_{3}}=0$ (so that $n_{_{3}} = m_{_{3}} = 0$). With this, the matrices ${\bf{R}}$ and ${\bf{S}}$ are nonzero only in their upper left $2 \times 2$ components. 

Let us now put $\tilde{W}_{_{13}} = \tilde{W}_{_{31}} = \tilde{W}_{_{23}} = \tilde{W}_{_{32}} = 0$, $\tilde{W}_{_{33}} = 1$, and consider an ansatz for the upper left $2 \times 2$ components of $\tilde{\bf{W}}$ such that:
\begin{eqnarray}
               \tilde{\bf{W}}_{_{(2 \times 2)}} = {{w - \tilde{w}_{_{1}}} \over {w - w_{_{1}}}}~
                                                  \bigl({{\bf{A}}}~(w- w_{_{1}}) + {{\bf{B}}} \bigr)~,  \label{Wtildevac}    
\end{eqnarray}
where the hermitian $2 \times 2$ matrices ${\bf{A}}$ and ${\bf{B}}$ are independent of $w$. With this, the constraint $(\ref{sec constr})$ is now trivially satisfied. Then, by comparing $(\ref{Wtildevac})$ with $(\ref{expWtilde})$, it can be seen that ${{\bf{A}}}$ and ${{\bf{B}}}$ are given by:
\begin{eqnarray}
                {{\bf{A}}} = 4 i~{\boldsymbol{\Omega}}_{_{(2 \times 2)}}~, 
                       \qquad \qquad
                {{\bf{B}}} = {\bf{W}}_{_{0}}(w_{_{1}})_{_{(2 \times 2)}} + 
                                               4 i~\bigl( \beta_{_{0}} - 
                                                         (w_{_{1}}-\tilde{w}_{_{1}}) 
                                                   \bigr )~{\boldsymbol{\Omega}}_{_{(2 \times 2)}} \label{calAB}
\end{eqnarray}
and, in place of $(\ref{sec constr})$, we obtain the alternative constraint that:  
\begin{eqnarray}
         &&   {\bf{R}}^\dagger~{\bf{W}}(w_{_{1}}) + {\bf{W}}(w_{_{1}})~{\bf{R}} + 4 i\,{\bf{R}}^\dagger~{\boldsymbol{\Omega}}~{\bf{R}}  \nonumber \\
         &&    \hspace{.8 in}
                = (w_{_{1}} - \tilde{w}_{_{1}})\, \Bigl(
                                                          {\bf{W}}(w_{_{1}}) +
                                                          4 i\,({\bf{R}}^\dagger~{\boldsymbol{\Omega}} + {\boldsymbol{\Omega}}~{\bf{R}})
                                                    \Bigr) -4 i\,(w_{_{1}}-\tilde{w}_{_{1}})^2~{\boldsymbol{\Omega}}~.    \label{second constr}
\end{eqnarray} 
This and $(\ref{first constr})$ set restrictions among the various parameters which enter the solution. 

Explicitly $\tilde{\bf{W}}$ is now given by:
\begin{eqnarray}
                  \tilde{\bf{W}}_{_{(2 \times 2)}} = {{w - \tilde{w}_{_{1}}} \over {w - w_{_{1}}}}~
                              \Bigl( {{\bf{W}}_{_{0}}}_{_{(2 \times 2)}} + 4 i \bigl( 
                                                                             \beta_{_{0}} - (w_{_{1}}- \tilde{w}_{_{1}})
                                                                          \bigl)~{\boldsymbol{\Omega}}_{_{(2 \times 2)}} \Bigr) ~,
    \qquad \tilde{W}_{_{33}} = 1~.  \label{tentWtildeelvac}
\end{eqnarray}
By simply requiring $h^{{12}}$ in $(\ref{ALEKbricks})$ to be real, we obtain $\beta_{_{0}} = w_{_{1}} - \tilde{w}_{_{1}}$. Consequently, ${\boldsymbol{\Psi}}_{_{0}}^\dagger~\tilde{\bf{W}}~{\boldsymbol{\Psi}}_{_{0}} \equiv \bf{K}$ assumes the form:
\begin{eqnarray}
               {\bf{K}}_{_{(2 \times 2)}} = {{w - \tilde{w}_{_{1}}} \over {w - w_{_{1}}}}~{{\bf{K}}_{_{0}}}_{_{(2 \times 2)}}~,
                    \qquad \qquad 
              K_{_{33}} = {K_{_{0}}}_{_{33}}~.   \label{KAPPA}
\end{eqnarray}
Since this is clearly a function of the spectral parameter $w$ only, the existence of solutions of the form $(\ref{1solRP})$ -- with the ansatz $(\ref{Wtildevac})$ -- is demonstrated in the vacuum case for any arbitrary seed. 

\smallskip

A simple electrovacuum generalization of the above method in which ${\bf A}$ and ${\bf B}$ are taken to be full $3 \times 3$ matrices is found to be inconsistent, except possibly for particular seed metrics. Among the reasons for this is the fact that $\beta_{_{0}} \neq w_{_{1}} - \tilde{w}_{_{1}}$ in this case.

\subsection{$N$-Soliton vacuum solutions}

The above approach may immediately be generalized to the case in which ${\boldsymbol{\chi}}$ has $N$ poles: i.e.:
\begin{subequations} \label{genNsolsol}
\begin{eqnarray}
             &&  \hspace{- .5 in}
                 {\boldsymbol{\chi}}     ~=~{\bf{I}}~+~\sum_{k = 1}^{N}~{1 \over{w -   w_{_{k}}}}~{\bf{R}}_{_{k}}    
                 \quad  \quad 
                 {\boldsymbol{\chi}}^{-1}~=~{\bf{I}}~+~\sum_{k = 1}^{N}~{1 \over{w - \tilde{w}_{_{k}}}}~{\bf{S}}_{_{k}}     \\
             &&  \hspace{- .5 in} 
                 {\bf{R}}_{_{k}} = {\bf{n}}_{_{k}} \otimes {\bf{m}}_{_{k}}, 
                 \qquad 
                 {\bf{S}}_{_{k}} = {\bf{p}}_{_{k}} \otimes {\bf{q}}_{_{k}},
                 \qquad 
                 w_{_{k}}, \tilde{w}_{_{k}} \in {\mathbb R},
                 \quad
                 w_{_{k}} \neq \tilde{w}_{_{k}}   \label{13.2bis} 
\end{eqnarray}
\end{subequations}
where ${\bf{n}}_{_{k}}$, ${\bf{m}}_{_{k}}$, ${\bf{p}}_{_{k}}$, ${\bf{q}}_{_{k}}$ are given in $(\ref{13.7})$ and $(\ref{13.14})$. Considering only the vacuum solutions, we must start with a vacuum seed and set the third components of the constant vectors ${\bf{k}}_{_{k}}$ and ${\bf{l}}_{_{k}}$ equal to zero (i.e. ${k_{_{k}}}_{3} = {l_{_{k}}}_{3} = 0$). Again, the proof of the existence of such a $N$-Soliton vacuum solution is based on a verification that ${\boldsymbol{\Psi}}^\dagger~{\bf{W}}~{\boldsymbol{\Psi}}$ is a function of the spectral parameter only, where:
\begin{eqnarray}
                {\bf{W}}~={\bf{W}}_{_{0}} - 4 i~( {\bf{R}}~{\boldsymbol{\Omega}}~+~{\bf{R}}^{\dagger}{\boldsymbol{\Omega}} )~+~4 i~\beta_{_{0}}~{\boldsymbol{\Omega}}~, 
                      \quad 
                {\bf{R}} =\sum_{k = 1}^N~{\bf{R}}_{_{k}}~, 
                      \quad 
                \beta_{_{0}} = \sum_{k = 1}^N~w_{_{k}} - \tilde{w}_{_{k}}.
\end{eqnarray} 
The function $\tilde{\bf{W}} = {\boldsymbol{\chi}}^\dagger~{\bf{W}}~{\boldsymbol{\chi}}$ is now given by:
\begin{eqnarray}
                && \tilde{\bf{W}}~=~{\bf{W}}~+~4 i \Bigl[
                                              {\bf{R}}^\dagger~{\boldsymbol{\Omega}}~+~{\boldsymbol{\Omega}}~{\bf{R}}
                                       \Bigl]~+~                                     \nonumber \\
                && \hspace{0.8 in}+ \sum_{k=1}^N~
                                   {1 \over {w - w_{_{k}}} }~
                                   \Bigl[
                                         {{\bf{R}}_{_{k}}}^\dagger~{\bf{W}}(w_{_{k}})~+~{\bf{W}}(w_{_{k}})~{\bf{R}}_{_{k}}
                                   \Bigl]~+~   \nonumber \\                   
                && \hspace{0.8 in}+ \sum_{k,j=1}^N~
                                   {1 \over {  (w - w_{_{k}}) ( w - w_{_{j}}) } }~
                                   {{\bf{R}}_{_{j}}}^\dagger~{\bf{W}}(w)~{\bf{R}}_{_{k}}~.             \label{expWtil2Nsol}
\end{eqnarray}
As in the $1$-Soliton case, it can be proved using $(\ref{important})$ that $\tilde{\bf{W}}$ contains simple poles only and there are $2 N$ constraints which can be used to set relations between the parameters of the solution. These are equivalent to $(\ref{first constr})$ and $(\ref{sec constr})$ and are given by: 
\begin{eqnarray}
                \overline{{\bf{n}}}_{_{k}} \cdot {\bf{W}}(w_{_{k}}) \cdot{\bf{n}}_{_{k}} ~=~0~, 
                      \qquad \qquad 
                \tilde{{\bf{W}}}(\tilde{w}_{_{k}}) \cdot{\bf{p}}_{_{k}} ~=~0~,
                       \qquad  \qquad  
                k = 1, \dots, N~. \label{first constr Nsol}
\end{eqnarray}

By adopting the ansatz $(\ref{Wtildevac})$ and using the procedure of adding one pole at a time, we find that the final expression for ${\bf{K}}$ is given by:
\begin{eqnarray}
                {{\bf{K}}}_{_{(2 \times 2)}} = \prod_{{k = 1}}^N
                                                        \left( { {w - \tilde{w}_{_{k}}} \over {w - w_{_{k}}} } \right)~
                                                        {{\bf{K}}_{_{0}}}_{_{(2 \times 2)}}~,
                    \qquad \qquad 
              K_{_{33}} = {K_{_{0}}}_{_{33}}~.   \label{KAPPANsol}
\end{eqnarray}
Since this depends on the spectral parameter only, the proof of the existence of vacuum soliton solutions in this class with $N$ real poles is demonstrated for any initial vacuum seed metric. 

\section{Explicit $1$-Soliton Solutions}   \label{sec:explsol}

For a generic vacuum seed metric, the corresponding matrix function ${\boldsymbol{\Psi}}_{_{0}}$ will have the form:
\begin{eqnarray}
                {\boldsymbol{\Psi}}_{_{0}}(w) = \left( \begin{array}{ccc}
                                                      \Psi_{_{11}}(w) & \Psi_{_{12}}(w) & 0 \\
                                                      \Psi_{_{21}}(w) & \Psi_{_{22}}(w) & 0 \\
                                                                 0 &            0 & 1
                                       \end{array} \right) \label{psi0}
\end{eqnarray}
To explicitly construct a general (electrovacuum) 1-soliton solution we may put:
\begin{eqnarray} 
                   {\bf{k}} = (1, k_{_{2}}, k_{_{3}} )
                    \qquad \quad
                   {\bf{l}} = (1, l_{_{2}}, l_{_{3}} )
\end{eqnarray} 
where the soliton suffix $\scriptstyle{1}$ has been suppressed. Then $(\ref{13.7})$ and $(\ref{13.14})$ take the form:
\begin{subequations} \label{miogenseed}
\begin{eqnarray}
             && {\bf{m}} = \left( X_{_{0}} ,  - Y_{_{0}} , k_{_{3}} \right)
                \qquad \qquad  \quad ~
                {\bf{p}} = \left( X_{_{\delta}},  Y_{_{\delta}}, l_{_{3}} \right) \\
             && \nonumber \\
             &&{\bf{q}} = - \delta~{{ \det {\boldsymbol{\Psi}}_{_{0}} (z_{_{0}}) } \over { \cal{D} }}~{\bf{m}}
                     \qquad \qquad 
                {\bf{n}} =   \delta~{{ \det {\boldsymbol{\Psi}}_{_{0}} (z_{_{0}}) } \over { \cal{D} }}~{\bf{p}}   
\end{eqnarray}
\end{subequations}
where $w_{_{1}} = z_{_{0}}$ and $\tilde{w}_{_{1}} = z_{_{0}} + \delta$, with $z_{_{0}}, \delta \in \mathbb R$, and the following quantities have been introduced:
\begin{subequations}
\begin{eqnarray} 
                && \hspace{-.6 in}
                   X_{_{0}}    =   {{{{\Psi}}_{_{22}}(z_{_{0}}) - k_{_{2}}~\Psi_{_{21}}(z_{_{0}})} 
                                           \over
                                     {\det {\boldsymbol{\Psi}}_{_{0}} (z_{_{0}})}}~
                            \qquad  \qquad  \quad
                   Y_{_{0}}    =   {{{{\Psi}}_{_{12}}(z_{_{0}}) - k_{_{2}}~\Psi_{_{11}}(z_{_{0}})} 
                                           \over 
                                     {\det {\boldsymbol{\Psi}}_{_{0}} (z_{_{0}})}}  \\
                && \nonumber \\
                && \hspace{-.6 in}
                   X_{_{\delta}} = \Psi_{_{11}}(z_{_{0}} + \delta) + l_{_{2}}~\Psi_{_{12}}(z_{_{0}} + \delta)
                            \qquad
                   Y_{_{\delta}} = \Psi_{_{21}}(z_{_{0}} + \delta) + l_{_{2}}~\Psi_{_{22}}(z_{_{0}} + \delta) \\
                && \nonumber \\
                && \hspace{-.6 in}
                   {\cal D} = X_{_{0}}~X_{_{\delta}} - Y_{_{0}}~Y_{_{\delta}} + k_{_{3}}~l_{_{3}} ~\det {\boldsymbol{\Psi}}_{_{0}} (z_{_{0}})
\end{eqnarray}
\end{subequations}
It is worth noticing here that each component of the vectors ${\bf{n}}$ and ${\bf{q}}$ is proportional to $\delta$. This confirms that a new solution can be obtained only if the poles in ${\boldsymbol{\chi}}$ and ${\boldsymbol{\chi}}^{-1}$ are distinct.

Let us consider the particular case of the Minkowski seed metric for which:
\begin{eqnarray} 
                {\bf{g}}_{_{0}} = \left( \begin{array}{cc} 
                                                - \epsilon & 0 \\
                                                0 & - \alpha^2
                           \end{array} \right),
                \quad
                {\boldsymbol{\Psi}}_{_{0}} = \left( \begin{array}{ccc} 
                                                            {1 \over {\sqrt{(w - \beta)^2 - \epsilon~\alpha^2}} } & 0      & 0      \\
                                                            i \epsilon {(w - \beta) \over {\sqrt{(w - \beta)^2 - \epsilon~\alpha^2}} } & 1 & 0      \\
                                                                                                   0 & 0      & 1
                                      \end{array} \right),  
                \quad
                {\bf{K}}_{_{0}} = \left( \begin{array}{ccc} 
                                                4 \epsilon & 0 & 0 \\
                                                0 & - 4 \epsilon & 0 \\
                                                0 &            0 & 1
                           \end{array} \right)   \label{minkowskiseed}
\end{eqnarray}
and construct the associated vacuum soliton solution with one real pole. We can therefore set $k_{_{3}} = l_{_{3}} = 0$. We can also choose $z_{_{0}} = 0$. By specializing $(\ref{miogenseed})$ to this seed, it can be shown that the constraints $(\ref{first constr})$ and $(\ref{second constr})$ give the following restrictions on the parameters:
\begin{eqnarray} 
                  | k_{_{2}} |^2 = 1~, \qquad \qquad | l_{_{2}} |^2 = 1~. 
\end{eqnarray}

\subsection{Vacuum Diagonal Solution}

To construct a diagonal solution, we can take:
\begin{eqnarray} 
                k_{_{2}} = i~,
                   \quad \quad ~
                l_{_{2}} = i~.
\end{eqnarray} 
With this, we obtain:
\begin{subequations} \label{tentativemetricminkd}
\begin{eqnarray} 
                && \hspace{-.5 in}
                   g_{_{11}} = \epsilon~
                               { {\delta + \epsilon \Bigl(
                                                          \sqrt{\beta^2 - \epsilon~\alpha^2} - 
                                                          \sqrt{(\beta - \delta)^2 - \epsilon~\alpha^2}
                                                    \Bigl)} 
                                   \over 
                                 {\delta - \epsilon \Bigl(
                                                          \sqrt{\beta^2 - \epsilon~\alpha^2} - 
                                                          \sqrt{(\beta - \delta)^2 - \epsilon~\alpha^2}
                                                    \Bigl)} } \\
                && \nonumber \\
                && \hspace{-.5 in}
                   g_{_{22}} = -\alpha^2 - 
                               2~\epsilon~\delta
                               { {\Bigl( \beta - \epsilon~\sqrt{\beta^2 - \epsilon~\alpha^2} \Bigl) 
                                  \Bigl( \beta - \delta - \epsilon~\sqrt{(\beta - \delta)^2 - \epsilon~\alpha^2} \Bigl)} 
                                   \over 
                                 {\delta - \epsilon~\Bigl(
                                                          \sqrt{\beta^2 - \epsilon~\alpha^2} - 
                                                          \sqrt{(\beta - \delta)^2 - \epsilon~\alpha^2}
                                                    \Bigl)} }   
\end{eqnarray}
\end{subequations} 
It is worth mentioning that the parameter $\beta_{_{0}}$ has been set as $\beta_{_{0}} = -~\delta$, according to the results in $\S~\ref{1solsolution}$. It can also be confirmed that $\det{\bf{g}} = \epsilon~\alpha^2$. Moreover the condition ${\boldsymbol{\Psi}}^\dagger {\bf{W}} {\boldsymbol{\Psi}} = {\bf{K}}(w)$ is fulfilled with:
\begin{eqnarray} 
                {\boldsymbol{\Psi}}^\dagger {\bf{W}} {\boldsymbol{\Psi}} = \left( \begin{array}{ccc} 
                                                                4~\epsilon~\Bigl(1 - {\delta \over w}\Bigl) & 0 & 0 \\
                                                                 0 & -4~\epsilon~\Bigl(1 - {\delta \over w}\Bigl) & 0 \\
                                                                 0 & 0 & 1
                                      \end{array} \right)   \label{KAPPAexpl}
\end{eqnarray}
which is in agreement with $(\ref{KAPPA})$. 

Let us now introduce the functions $\mu_{_{1}}^{\pm}$ and $\mu_{_{2}}^{\pm}$ given by
\begin{eqnarray} 
               \mu_{_{1}}^{\pm} = - \beta \pm \sqrt{\beta^2 - \epsilon \alpha^2}~,
                       \qquad \qquad
               \mu_{_{2}}^{\pm} = - (\beta - \delta) \pm \sqrt{(\beta - \delta)^2 - \epsilon \alpha^2}~,   \label{pol-traj}
\end{eqnarray}
which are the so-called pole-trajectories that are typical in the BZ formalism. With these, the above metric $(\ref{tentativemetricminkd})$ can be rewritten in the form:
\begin{subequations}  \label{MY2RPd}
\begin{eqnarray} 
    && \hspace{-.4 in} 
       \epsilon~= +1: \qquad g_{_{11}} = { {\mu_{_{1}}^{-}~-~\mu_{_{2}}^{-}} 
                                           \over
                                          {\mu_{_{1}}^{+}~-~\mu_{_{2}}^{+}}} = 
                                        { { {{\alpha^2} \over {\mu_{_{1}}^{+}}}~-~{{\alpha^2} \over {\mu_{_{2}}^{+}}} } 
                                           \over
                                          {\mu_{_{1}}^{+}~-~\mu_{_{2}}^{+}}}= 
                             - \> {\alpha^2 \over { \mu_{_{1}}^{+}~{\mu}_{_{2}}^{+} } }~, 
                            \hspace{.7 in}
                            g_{_{22}} = {\alpha^2 \over {g_{_{11}}}}  \\
    && \hspace{-.4 in} 
       \epsilon~= -1: \qquad g_{_{11}} = -{ {\mu_{_{1}}^{+}~-~\mu_{_{2}}^{+}} 
                                           \over
                                          {\mu_{_{1}}^{-}~-~\mu_{_{2}}^{-}}} = 
                                        -{ { {{-\alpha^2} \over {\mu_{_{1}}^{-}}}~-~{{-\alpha^2} \over {\mu_{_{2}}^{-}}} } 
                                           \over
                                          {\mu_{_{1}}^{-}~-~\mu_{_{2}}^{-}}}= 
                             - \> {\alpha^2 \over { \mu_{_{1}}^{-}~{\mu}_{_{2}}^{-} } }~,
                            \hspace{.3 in}
                            g_{_{22}} = - \>{\alpha^2 \over {g_{_{11}}}}
\end{eqnarray} 
\end{subequations} 
It may immediately be observed that this solution is identical to the BZ soliton solution with two real poles and the same seed $\cite{CAV1}$. (The global structure of this solution is discussed in \cite{GRpap2}).

\subsection{Vacuum Non-diagonal Solution}

Let us consider the following parameters:
\begin{eqnarray} 
                k_{_{2}} = 1~,
                     \qquad \qquad 
                l_{_{2}} = 1  ~.
\end{eqnarray}
This choice generates a nondiagonal solution given by: 
\begin{eqnarray} 
                && \hspace{-.3 in}
                   g_{_{11}} = -~\epsilon~
                                { {  \beta^2 - \beta~\delta - \epsilon~\alpha^2 + 
                                     \sqrt{\beta^2 - \epsilon~\alpha^2}~\sqrt{(\beta-\delta)^2 - \epsilon~\alpha^2} } 
                                   \over
                                  {  \beta^2 - \beta~\delta + \delta^2 - \epsilon~\alpha^2 + 
                                     \sqrt{\beta^2 - \epsilon~\alpha^2}~\sqrt{(\beta-\delta)^2 - \epsilon~\alpha^2} } }     \nonumber \\
                && \nonumber \\
                && \hspace{-.3 in}
                   g_{_{12}} =  \delta~\epsilon~
                                { {\sqrt{\beta^2 - \epsilon~\alpha^2}~(\beta - \delta)~+~\beta \sqrt{(\beta-\delta)^2 - \epsilon~\alpha^2}} 
                                   \over
                                  {  \beta^2 - \beta~\delta + \delta^2 - \epsilon~\alpha^2 + 
                                     \sqrt{\beta^2 - \epsilon~\alpha^2}~\sqrt{(\beta-\delta)^2 - \epsilon~\alpha^2} } }   \label{tentativemetricmink}\\
                && \nonumber \\
                && \hspace{-.3 in}
                   g_{_{22}} =  
                                -{ {- \epsilon~\alpha^4 + 2~\epsilon~\beta~(\beta - \delta)~\delta^2 +
                                   \alpha^2~\bigl( 
                                                  \beta^2 - \beta~\delta + 2 \delta^2 + 
                                                  \sqrt{\beta^2 - \epsilon~\alpha^2}~\sqrt{(\beta-\delta)^2 - \epsilon~\alpha^2}
                                            \bigl)}   
                                   \over
                                  {  \beta^2 - \beta~\delta + \delta^2 - \epsilon~\alpha^2 + 
                                     \sqrt{\beta^2 - \epsilon~\alpha^2}~\sqrt{(\beta-\delta)^2 - \epsilon~\alpha^2} } }  \nonumber
\end{eqnarray} 
It can again be confirmed that $\det{\bf{g}} = \epsilon~\alpha^2$ and that the condition ${\boldsymbol{\Psi}}^\dagger {\bf{W}} {\boldsymbol{\Psi}} = {\bf{K}}(w)$ is fulfilled. Indeed, it is found that ${\bf{K}}(w)$ is identical to that for the diagonal case $(\ref{tentativemetricminkd})$ -- see formula $(\ref{KAPPAexpl})$. This is inevitable from the fact that the formula $(\ref{KAPPA})$ does not contain the free parameters ${\bf{k}}_{_{i}}$ and ${\bf{l}}_{_{i}}$ -- it only contains the parameter $\delta$ that appears in the poles.

Using the pole-trajectories $(\ref{pol-traj})$, the above metric components can be rewritten in the form:
\begin{eqnarray} 
               g_{_{11}} = - 2 \epsilon~{ { (\mu_{_{1}}^{+} - \mu_{_{2}}^{-} )~
                                                (\mu_{_{2}}^{+} - \mu_{_{1}}^{-} ) }
                                               \over
                                              { (\mu_{_{1}}^{+} - \mu_{_{2}}^{-} )^2 + 
                                                (\mu_{_{2}}^{+} - \mu_{_{1}}^{-} )^2 } }~, 
                  \qquad \qquad
               g_{_{12}} = -2 \epsilon~\delta~{ { \mu_{_{1}}^{+} \mu_{_{2}}^{+} - 
                                                      \mu_{_{1}}^{-} \mu_{_{2}}^{-} }
                                                          \over
                                                    { (\mu_{_{1}}^{+} - \mu_{_{2}}^{-} )^2 + 
                                                      (\mu_{_{2}}^{+} - \mu_{_{1}}^{-} )^2 } }~.   \label{MY2RPnd}
\end{eqnarray}
The $g_{_{22}}$ component can be easily obtained by way of the usual condition that $\det {\bf{g}} = \epsilon \, \alpha^2$.  

\section{Conclusions}

Generally, the Alekseev $N$-soliton method for the construction of new
solutions of the Einstein--Maxwell equations requires the addition of $N$ distinct 
(normally complex conjugate) poles in the inverse of the scattering matrix. In the vacuum case, the method
is equivalent to that of Belinskii and Zakharov (with $2N$ poles) provided
the poles occur as complex conjugate pairs.

The purpose of this paper has been to modify the Alekseev inverse-scattering
method to permit the use of real poles. This has been attempted by
introducing distinct real poles in the inverse matrix. We have found that
this construction is successful at least in the vacuum case in which it has shown to be equivalent to the BZ method 
with distinct real poles.

\section*{Acknowledgments}

The authors are greatly indebted to Dr G. A. Alekseev for extensive
discussions on these topics and for his comments on an early version of this paper.


\end{document}